\documentclass[notitlepage,prd,a4paper,11pt,amsfonts,amssymb,amsmath]{revtex4}
\usepackage{enumerate}

\newcommand{\mat}[1]{\underline{\underline{#1}}}
\newcommand{\vek}[1]{\underline{#1}}

\begin{document}
\title{Multimetric extension of the PPN formalism: experimental\\consistency of repulsive gravity}

\author{Manuel Hohmann}
\email{manuel.hohmann@desy.de}
\affiliation{Zentrum f\"ur Mathematische Physik und II. Institut f\"ur Theoretische Physik, Universit\"at Hamburg, Luruper Chaussee 149, 22761 Hamburg, Germany}
\author{Mattias N.\,R. Wohlfarth}
\email{mattias.wohlfarth@desy.de}
\affiliation{Zentrum f\"ur Mathematische Physik und II. Institut f\"ur Theoretische Physik, Universit\"at Hamburg, Luruper Chaussee 149, 22761 Hamburg, Germany}

\begin{abstract}
Recently we discussed a multimetric gravity theory containing several copies of standard model matter each of which couples to its own metric tensor. This construction contained dark matter sectors interacting repulsively with the visible matter sector, and was shown to lead to cosmological late time acceleration. In order to test the theory with high-precision experiments within the solar system we here construct a simple extension of the parametrized post-Newtonian (PPN) formalism for multimetric gravitational backgrounds. We show that a simplified version of this extended formalism allows the computation of a subset of the PPN parameters from the linearized field equations. Applying the simplified formalism we find that the PPN parameters of our theory do not agree with the observed values, but we are able to improve the theory so that it becomes consistent with experiments of post-Newtonian gravity and still features its promising cosmological properties.
\end{abstract}
\maketitle

\section{Introduction}\label{sec:motivation}
In a recent article~\cite{Hohmann:2010vt} we presented a multimetric gravity theory containing \(N > 2\) copies $\Psi^1\dots\Psi^N$ of standard model matter and a corresponding number of metric tensors $g^1\dots g^N$. The action was of the type
\begin{equation}\label{eqn:actionsplit}
S = S_G[g^1, \ldots, g^N] + \sum_{I = 1}^{N}S_M[g^I, \Psi^I]\,.
\end{equation}
This structure tells us that the interaction between the different standard model copies with action $S_M$ is mediated only by the coupling of the metrics in the gravitational action~$S_G$. Hence particles contained in one standard model sector interact with particles from a different sector through an extremely small cross section involving the squared Newton's constant. In the astronomical context, each type of matter appears dark to observers residing in a different sector. In the Newtonian limit, the gravitational interaction was constructed to be attractive within each sector and repulsive of equal strength between different sectors (which was shown in~\cite{Hohmann:2009bi} not to be possible in the bimetric case $N=2$). We derived a simple cosmological model and could show that, due to the repulsion between different types of matter, it naturally features an accelerating late-time expansion of the universe.

A wide range of other alternative gravity theories have been proposed in order to model the observed late-time cosmology. These include: modified Newtonian dynamics~\cite{Milgrom:1983ca} which is not a geometric theory; theories which in addition to a metric contain scalar and vector fields to describe gravitational backgrounds~\cite{Bekenstein:2004ne,Bekenstein:2004ca}; curvature corrections in metric theories which may appear as full Riemann tensor corrections as in~\cite{Schuller:2004nn}, or as Ricci scalar corrections in $f(R)$ theories~\cite{Sotiriou:2008rp}; higher-dimensional models such as~\cite{Dvali:2000hr,Lue:2005ya}; and structural extensions such as non-symmetric gravity theory~\cite{Moffat:1995fc} and area metric gravity~\cite{Punzi:2006hy,Punzi:2006nx}. Of course all alternative theories of gravity have to be tested against the data available from solar system experiments. For the alternative theories mentioned above this issue has been discussed, for example, in~\cite{DeFelice:2009aj,Ruggiero:2008zj,Battat:2008bu,Chiba:2006jp,Faraoni:2006hx,Punzi:2008dv}.

It is the aim of this article to further test the predictions of our multimetric theory by high-precision solar system experiments. For this purpose we will extend the parametrized post-Newtonian (PPN) formalism to multimetric gravity and then apply it to our theory. This elaborate formalism was developed mainly by Nordtvedt~\cite{Nordtvedt:1968qs} and Will~\cite{Thorne:1970wv} to test single metric gravity theories, see~\cite{Will:1993ns} for a review. It assigns to each gravity theory a set of ten experimentally measurable quantities, the so-called PPN parameters \(\beta, \gamma, \xi, \alpha_1\dots \alpha_3, \zeta_1\dots\zeta_4\). These parameters appear as coefficients in a perturbative expansion of the metric tensor and can be computed by a perturbative solution of the equations of motion. The values of the PPN parameters of a theory are closely linked to its physical properties. Most notably, they measure the non-linearity in the Newtonian superposition law for gravity, the spatial curvature generated by matter sources, preferred frame effects and the failure of conservation of energy, momentum and angular momentum.

Most PPN parameters have been determined by a wide range of high precision experiments. Their values are fixed within very narrow bounds at \(\beta = \gamma = 1\) while all other parameters vanish~\cite{WillReview}. This means that there is no experimental evidence for peferred frame effects or a failure of conservation of energy, momentum or angular momentum. In particular, the bounds are: \(|\gamma - 1| < 2.3 \cdot 10^{-5}\) from Cassini tracking; \(|\beta - 1| < 3 \cdot 10^{-3}\) from helioseismology; \(|\xi| < 10^{-3}\) from gravimeter data of the Earth tides; \(|\alpha_1| < 10^{-4}\) from lunar laser ranging; \(|\alpha_2| < 4 \cdot 10^{-7}\) from the solar alignment with the ecliptic; \(|\alpha_3| < 4 \cdot 10^{-20}\) from pulsar statistics; \(|\zeta_2| < 4 \cdot 10^{-5}\) from observations of the binary pulsar PSR 1913+16; \(|\zeta_3| < 10^{-8}\) from lunar acceleration; \(|\zeta_1| < 2 \cdot 10^{-2}\) and \(|\zeta_2| < 6 \cdot 10^{-3}\) from combinations of the other PPN measurements. Therefore theories for which the PPN parameters take significantly different values are experimentally excluded.

In this article we will show that these high-precision tests can also be used to test multimetric gravity theories where the gravitational field is described by several metric tensors each of which governs the motion of one type of matter and a corresponding class of observers. In section~\ref{sec:multippn} we will construct a simple extension of the parametrized post-Newtonian formalism to multimetric gravity. We consider the physically relevant situation of sources formed from a single type of matter for which we show that each multimetric gravity theory is characterized by an extended set of post-Newtonian parameters, in complete analogy to the standard PPN formalism. We will then show in section~\ref{sec:linearppn} that a subset of these PPN parameters can already be obtained from the linearized field equations. In section~\ref{sec:application} we finally apply the linearized multimetric formalism to our particular gravity model for the accelerating late universe~\cite{Hohmann:2010vt} and explicitly compute its PPN parameters. We find that these do not match the observed values, but we are able to present an improved theory from which the correct values are obtained while the cosmological results are unchanged. We will conlude with a discussion in section~\ref{sec:conclusion}. The appendix contains some important details of the computation.

\section{Multimetric extension of the PPN formalism}\label{sec:multippn}
The PPN formalism was originally developed for single metric gravity theories, see the review in~\cite{Will:1993ns}. In this section we present a simple extension of this formalism for multimetric gravity theories. This extension is contructed to describe the physical situation of the solar system for which we will argue that two of the $N$ metrics suffice. These two metric tensors will be expressed in terms of the PPN potentials and an extended set of PPN parameters. Comparison of these with the standard PPN parameters then enables tests of multimetric gravity by high-precision data.

The class \eqref{eqn:actionsplit} of multimetric gravity theories we consider is restricted by the following four assumptions which were motivated in full detail in~\cite{Hohmann:2009bi,Hohmann:2010vt}:
\begin{enumerate}[\it (i)]
\item\label{ass:tensor}
The field equations are obtained by variation with respect to the metrics $g^1_{ab}\dots g^N_{ab}$, and so are a set of symmetric two-tensor equations of the form \(\vek{K_{ab}} = 8\pi G_N\vek{T_{ab}}\).

\item\label{ass:derivatives}
The geometry tensor \(\vek{K_{ab}}\) contains at most second derivatives of the metric, which can be achieved by a suitable choice of the gravitational action.

\item\label{ass:symmetry}
The field equations are symmetric with respect to arbitrary permutations of the sectors $(g^I,\Psi^I)$, which can be understood as a generalized Copernican principle.

\item\label{ass:flat}
The vacuum solution is given by a set of flat metrics \(g^I_{ab} = \eta_{ab}\). (Poincar\'e symmetry for all metrics simultaneously implies \(g^I_{ab} = \lambda^I\eta_{ab}\) for constants $\lambda^I$ that, invoking the Copernican principle for the vacuum, should be equal and can be set to $\lambda^I=1$.)
\end{enumerate}
Cosmological constants are excluded because we are interested in multimetric gravity theories in which the accelerating universe is modelled by a repulsive interaction between different standard model copies, as we have seen in~\cite{Hohmann:2010vt}.

Another consequence of the repulsion of matter from different sectors is their separation as the universe evolves. Hence we may safely assume that the gravitational field in regions like our solar system is dominated by a single type of matter; we formulate:
\begin{enumerate}[\it (i)]
\setcounter{enumi}{4}
\item\label{ass:singlesource}Regions exist where the gravitational field is generated by matter sources from a single sector.
\end{enumerate}
Combining this assumption with the symmetry assumption \textit{(\ref{ass:symmetry})} guarantees the existence of solutions in which the metric and the energy-momentum tensor corresponding to the dominant matter source are distinct, while all other metric tensors are equal and their energy-momentum tensors vanish. We assume that this simplest solution is actually realized, since, as observers in a distinct region, we have no further detailed access to the physics of the other sectors:
\begin{enumerate}[\it (i)]
\setcounter{enumi}{5}
\item\label{ass:equalsol}The metric tensors of all other sectors besides the one distinguished by the dominant matter source are equal.
\end{enumerate}
In the following we will indicate all quantities within the distinct sector \(I = 1\) by a superscript `$+$', and all quantities within the other sectors by a superscript `$-$'. Hence $g^+_{ab}=g^1_{ab}$ and $T^+_{ab}=T^1_{ab}$ while $g^-_{ab}=g^2_{ab}=\dots=g^N_{ab}$ and \(T^-_{ab} = T^2_{ab} = \dots = T^N_{ab}=0\). We will now extend the standard single metric PPN formalism to this physical situation.

Basic ingredient of the PPN formalism is an expansion of the geometric background in orders of the velocity of the source matter. Using assumption \textit{(\ref{ass:flat})} this is a weak field approximation around the flat vacuum metric $\eta$ in Cartesian coordinates $(x^0,x^\alpha)$,
\begin{equation}\label{eqn:linan}
g^{\pm}_{ab} = \eta_{ab} + h^{\pm}_{ab} =\eta_{ab}+ h^{(1)\pm}_{ab}+ h^{(2)\pm}_{ab}+ h^{(3)\pm}_{ab}+ h^{(4)\pm}_{ab}\,.
\end{equation}
Higher than fourth velocity order $\mathcal{O}(4)$ is not considered. It turns out that not all metric perturbations $h^{(i)\pm}_{ab}\sim\mathcal{O}(i)$ are relevant to describe the motion of test bodies. Moreover certain components vanish due to Newtonian energy conservation or time reversal symmetry. We now only list the relevant, non-vanishing components of the metric perturbations. These are written in terms of the so-called PPN potentials $U,U_{\alpha\beta},V_{\alpha}, W_{\alpha},\Phi_W, \Phi_1\dots \Phi_4, \mathcal{A}, \mathcal{B}$ and constant PPN parameters~$\alpha^\pm,\gamma^\pm,\theta^\pm,\sigma^\pm_\pm,\beta^\pm,\xi^\pm,\phi^\pm_1\dots\phi^\pm_4,\mu^\pm,\nu^\pm$ as
\begin{subequations}\label{eqn:linan2}
\begin{align}
h^{(2)\pm}_{00} &= 2\alpha^{\pm}U\,,\\
h^{(2)\pm}_{\alpha\beta} &= 2\gamma^{\pm}U\delta_{\alpha\beta} + 2\theta^{\pm}U_{\alpha\beta}\,,\\
h^{(3)\pm}_{0\alpha} &= \sigma^{\pm}_+(V_{\alpha} + W_{\alpha}) + \sigma^{\pm}_-(V_{\alpha} - W_{\alpha})\,,\\
h^{(4)\pm}_{00} &=- 2\beta^{\pm}U^2 - 2\xi^{\pm}\Phi_W + 2\phi^{\pm}_1\Phi_1 + 2\phi^{\pm}_2\Phi_2 + 2\phi^{\pm}_3\Phi_3 + 2\phi^{\pm}_4\Phi_4 + 2\mu^{\pm}\mathcal{A} + 2\nu^{\pm}\mathcal{B}\,.
\end{align}
\end{subequations}
The spacetime dependent PPN potentials are Poisson-like integrals over the source matter energy density \(\rho\), velocity \(v^{\alpha}\), internal energy \(\rho\Pi\) and pressure \(p\). Due to the virial theorem, the energy density is associated a velocity order $\rho\sim\mathcal{O}(2)$ while one assigns $\rho\Pi,p\sim\mathcal{O}(4)$. The PPN potentials at second velocity order are the standard Newtonian potential $U$ and a tensor potential $U_{\alpha\beta}$,
\begin{equation}
U(x^0,\vec{x}) = \int\mathrm{d}^3x'\frac{\rho(x^0,\vec{x}')}{|\vec{x} - \vec{x}'|}\,, \qquad U_{\alpha\beta}(x^0,\vec{x}) = \int\mathrm{d}^3x'\frac{\rho(x^0,\vec{x}')(x_{\alpha} - x_{\alpha}')(x_{\beta} - x_{\beta}')}{|\vec{x} - \vec{x}'|^3}\,.
\end{equation}
Similar integrals define the third order vector potentials
\begin{subequations}\label{eqn:VWdef}
\begin{eqnarray}
V_{\alpha}(x^0,\vec{x}) &=& \int\mathrm{d}^3x'\frac{\rho(x^0,\vec{x}')v_\alpha(x^0,\vec{x}')}{|\vec{x} - \vec{x}'|}\,,\\
W_{\alpha}(x^0,\vec{x}) &=& \int\mathrm{d}^3x'\frac{\rho(x^0,\vec{x}')v_\beta(x^0,\vec{x}')(x_{\alpha} - x_{\alpha}')(x_{\beta} - x_{\beta}')}{|\vec{x} - \vec{x}'|^3}
\end{eqnarray}
\end{subequations}
and the fourth order scalar potentials \(\Phi_W, \Phi_1, \Phi_2, \Phi_3, \Phi_4, \mathcal{A}, \mathcal{B}\), see~\cite{Will:1993ns} for full detail.

The metric ansatz presented above fully reduces to the PPN formalism in single metric theories. To see this, one simply drops all superscripts `$\pm$'. Moreover, it is then conventional to choose a gauge so that the parameters $\theta$ and $\nu$ vanish, thus removing the potentials $U_{\alpha\beta}$ and $\mathcal{B}$. Also, one may choose $\alpha=1$ by absorbing its value into the definition of the gravitational constant. In the extended PPN formalism discussed here we have the same amount of freedom. However, gauge choices result from diffeomorphism invariance and affect all metrics simultaneously, and also a rescaling of the gravitational constant in one sector will affect the gravitational interaction between all sectors. Hence only one of the two metrics $g^\pm$ can be simplified.

Since we wish to compare the multimetric PPN parameters to experimental data, we turn our focus to observers that reside within the distinct sector, i.e., to observers for whom the dominating matter source is visible. These observers as well as their visible type of matter are affected by the metric \(g^+\) only. Thus \(g^+\) corresponds to the single metric \(g\) in the standard PPN formalism. We will therefore choose a gauge in which \(h^+\) has the standard PPN form, i.e., \(\theta^+ = \nu^+ = 0\), and fix the gravitational constant so that \(\alpha^+ = 1\). The remaining ten PPN parameters contained in \(h^+\) are then identified with the PPN parameters known from the single metric case by direct comparison. The conversion between the standard PPN parameters used in~\cite{Will:1993ns} and the notation we use in the metric ansatz~\eqref{eqn:linan2} is given by the relations
\begin{gather}
\beta^+ = \beta\,, \quad \gamma^+ = \gamma\,, \quad \xi^+ = \xi\,, \quad \phi^+_1 = \frac{1}{2}(2 + 2\gamma + \alpha_3 + \zeta_1 - 2\xi)\,,\nonumber\\
\phi^+_2 = 1 + 3\gamma - 2\beta + \zeta_2 + \xi\,, \quad \phi^+_3 = 1 + \zeta_3\,, \quad \phi^+_4 = 3\gamma + 3\zeta_4 - 2\xi\,,\\
\mu^+ = \frac{1}{2}(2\xi - \zeta_1)\,, \quad \sigma^+_+ = -\frac{1}{4}(4 + 4\gamma + \alpha_1)\,, \quad \sigma^+_- = -\frac{1}{4}(2 + 4\gamma + \alpha_1 - 2\alpha_2 + 2\zeta_1 - 4\xi)\,.\nonumber
\end{gather}
With this identification we can convert the experimentally measured values of the standard PPN parameters to our notation and obtain
\begin{equation}\label{eqn:ppnvalues}
\beta^+ = \gamma^+ = \phi^+_3 = 1, \quad \xi^+ = \mu^+ = 0, \quad \phi^+_1 = \phi^+_2 = 2, \quad \phi^+_4 = 3, \quad \sigma^+_+ = -2, \quad \sigma^+_- = -\frac{3}{2}\,.
\end{equation}
As we already argued in the introduction, these values are fixed by numerous experiments within very narrow bounds. Gravity theories with significantly different PPN parameter values are therefore experimentally excluded.

The values of the extended PPN parameters can now be computed in complete analogy to the standard PPN formalism. We will examine this procedure in the following section~\ref{sec:linearppn}. Conveniently, it will turn out that some of the extended PPN parameters can already be obtained from the linearized equations of motion.

\section{Linearized PPN formalism}\label{sec:linearppn}
In the preceding section we have discussed an extension of the PPN framework to multimetric gravity theories, which allows tests of these theories by means of available experimental data from solar system and astronomical experiments. We have shown that, in addition to the parameters obtained from the standard PPN framework, we obtain further parameters that characterize the influence of matter sources from one sector on the metric tensors of the other sectors. In this section, we will show that a number of the extended PPN parameters can be computed already from the linearized equations of motion. This covers the PPN velocity orders up to $\mathcal{O}(3)$ while all terms of $\mathcal{O}(4)$ and higher are neglected. We will give a step-by-step recipe for this calculation.

\subsection{Geometry and matter content}\label{subsec:geometry}
The starting point for our computation is the most general linearized field equations compatible with our assumptions stated at the beginning of section~\ref{sec:multippn}. Using the convention \(G_N = 1\) for the normalization of Newton's constant, these equations read
\begin{equation}\label{eqn:ppnlineq}
\vek{K}_{ab} = 8\pi\vek{T}_{ab}
\end{equation}
with the linearized geometry tensor
\begin{equation}\label{eqn:ppngeom}
\vek{K}_{ab} = \mat{P} \cdot \partial^p\partial_{(a}\vek{h}_{b)p} + \mat{Q} \cdot \square\vek{h}_{ab} + \mat{R} \cdot \partial_a\partial_b\vek{h} + \mat{M} \cdot \partial_p\partial_q\vek{h}^{pq}\eta_{ab} + \mat{N} \cdot \square\vek{h}\eta_{ab}+\mathcal{O}(h^2)
\end{equation}
and constant parameter matrices $P,Q,R,M,N$. Our assumptions pose two restrictions on these matrices. First, the linearized equations of motion \eqref{eqn:ppnlineq} are invariant under a permutation of the sectors by assumption \textit{(\ref{ass:symmetry})}. Hence the parameter matrices will have the form
\begin{equation}\label{eqn:parmatsym}
O^{IJ} = O^- + (O^+ - O^-)\delta^{IJ}
\end{equation}
with diagonal entries \(O^+\) and off-diagonal entries \(O^-\) for \(O = P, Q, R, M, N\). This leaves us with a set of ten parameters determined by the underlying (nonlinear) gravity theory. Second, the equations of motion are tensor equations by assumption \textit{(\ref{ass:tensor})}, and so the linearized equations should be gauge-invariant. Using the formalism of gauge-invariant perturbation theory detailed in~\cite{Hohmann:2009bi}, one finds the invariance conditions
\begin{equation}\label{eqn:gaugeinv1}
(\mat{P} + 2\mat{Q}) \cdot \vek{1} = (\mat{P} + 2\mat{R}) \cdot \vek{1} = (\mat{M} + \mat{N}) \cdot \vek{1} = \vek{0}\,,
\end{equation}
where \(\vek{1}=(1,\dots,1)^t\) and \(\vek{0}=(0,\dots,0)^t\) denote $N$-component vectors. Using \eqref{eqn:parmatsym}, these conditions can be written in the form
\begin{subequations}\label{eqn:gaugeinv2}
\begin{align}
P^+ + (N - 1)P^- + 2Q^+ + 2(N - 1)Q^- &= 0\,,\\
P^+ + (N - 1)P^- + 2R^+ + 2(N - 1)R^- &= 0\,,\\
M^+ + (N - 1)M^- + N^+ + (N - 1)N^- &= 0\,.
\end{align}
\end{subequations}
These equations fix three of the ten parameters in the parameter matrices so that the most general linearized curvature tensor consistent with our assumptions is completely determined by a set of seven parameters.

We will now turn our attention from the geometry side to the matter side of the equations of motion. Recall that, according to assumption \textit{(\ref{ass:singlesource})}, we consider only solutions of the field equations in which the gravitational field is generated by matter sources within a single sector, i.e., by a single energy-momentum tensor \(T^+\), while all other energy-momentum tensors \(T^-\) must vanish. In order to solve the linearized field equations~\eqref{eqn:ppnlineq}, we need to expand \(T^+\) up to the required order of perturbation theory, i.e., to velocity order $\mathcal{O}(3)$. We will use the ansatz
\begin{equation}\label{eqn:ppnlinmat}
T^+_{00} = \rho\,, \quad T^+_{0\alpha} = -\rho v_{\alpha}\,, \quad T^+_{\alpha\beta} = 0\,,
\end{equation}
corresponding to a perfect fluid of density \(\rho\sim\mathcal{O}(2)\) and velocity \(v^{\alpha}\sim\mathcal{O}(1)\).

\subsection{Computation of the extended PPN parameters}\label{subsec:ppncomp}
We will now explicitly solve the equations of motion. Omitting all terms in the PPN metric~\eqref{eqn:linan2} corresponding to perturbations of velocity order $\mathcal{O}(4)$ we may use the simplified ansatz
\begin{subequations}\label{eqn:ppnlinmet}
\begin{align}
h^{\pm}_{00} &= -\alpha^{\pm}\triangle\chi\,,\\
h^{\pm}_{0\alpha} &= \sigma^{\pm}_+X^+_{\alpha} + \sigma^{\pm}_-X^-_{\alpha}\,,\\
h^{\pm}_{\alpha\beta} &= 2\theta^{\pm}\partial_\alpha\partial_\beta\chi - (\gamma^{\pm} + \theta^{\pm})\triangle\chi\delta_{\alpha\beta}\,.
\end{align}
\end{subequations}
These expressions are rewritten in terms of the so-called superpotential
\begin{equation}
\chi(x^0,\vec{x}) = -\int\mathrm{d}^3x'\rho(x^0,\vec{x}')|\vec{x} - \vec{x}'|
\end{equation}
using the relations
\begin{equation}
U = -\frac{1}{2}\triangle\chi\,, \quad U_{\alpha\beta} = \partial_\alpha\partial_\beta\chi - \frac{1}{2}\triangle\chi\delta_{\alpha\beta}\,.
\end{equation}
We have furthermore introduced the notation \(X^{\pm}_{\alpha} = V_{\alpha} \pm W_{\alpha}\) for the vector potentials. The advantage in using \(X^{\pm}_{\alpha}\) instead of \(V_{\alpha}\) and \(W_{\alpha}\) results from the fact that \(X^-_{\alpha} = \partial_\alpha\partial_0\chi\) is a pure divergence and \(X^+_{\alpha}\) is a divergence-free vector, \(\partial^\alpha X^+_\alpha = 0\). These relations follow from the Newtonian continuity equation $\partial_0\rho + \partial_\alpha (\rho v^\alpha)=0$ and the definitions~\eqref{eqn:VWdef}, and will be used repeatedly in the following computation.

We begin by performing a $(1 + 3)$-split of the equations of motion \eqref{eqn:ppnlineq}. Using the energy-momentum tensor ansatz~\eqref{eqn:ppnlinmat} we obtain the equations
\begin{subequations}\label{eqn:ppnlin}
\begin{align}
K^{+}_{00} &= 8\pi\rho\,, & K^{-}_{00} &= 0\,,\label{eqn:ppnlinscal}\\
K^{+}_{0\alpha} &= -8\pi\rho v_{\alpha}\,, & K^{-}_{0\alpha} &= 0\,,\label{eqn:ppnlinvect}\\
K^{+}_{\alpha\beta} &= 0\,, & K^{-}_{\alpha\beta} &= 0\,.\label{eqn:ppnlintens}
\end{align}
\end{subequations}
In order to solve these equations, we expand the geometry tensor $\vek K_{ab}$ given in~\eqref{eqn:ppngeom} using the PPN metric \eqref{eqn:ppnlinmet}. In this calculation we once again drop all terms of velocity order $\mathcal{O}(4)$, taking care of the fact that time derivatives count as $\partial_0\sim\mathcal{O}(1)$. Up to the required order $\mathcal{O}(3)$ the geometry tensor then takes the form
\begin{subequations}\label{eqn:ppnlincoeff}
\begin{align}
K^{\pm}_{00} &= c^{\pm}_1\triangle\triangle\chi\,,\\
K^{\pm}_{0\alpha} &= c^{\pm}_2\triangle X^+_{\alpha} + c^{\pm}_3\triangle X^-_{\alpha}\,,\\
K^{\pm}_{\alpha\beta} &= c^{\pm}_4\triangle\triangle\chi\delta_{\alpha\beta} + c^{\pm}_5\triangle\partial_\alpha\partial_\beta\chi\,,
\end{align}
\end{subequations}
where the coefficients \(c^{\pm}_1, \ldots, c^{\pm}_5\) are constants which depend linearly both on the PPN parameters and the components of the parameter matrices \eqref{eqn:parmatsym}. For a detailed expansion of these coefficients, see appendix \ref{sec:ppnlincoeff}.

We will now determine the coefficients \(c^{\pm}_1, \ldots, c^{\pm}_5\) such that the equations of motion are satisfied for arbitrary matter distributions $\rho$ and $v^\alpha$. First, we solve the scalar equations \eqref{eqn:ppnlinscal}. Using the relation
\begin{equation}
\triangle\triangle\chi = -2\triangle U = 8\pi\rho\,,
\end{equation}
one can see that these are solved if, and only if, the corresponding coefficients take the values
\begin{equation}\label{eqn:12}
c^+_1 = 1\,, \quad c^-_1 = 0\,.
\end{equation}
We continue with the tensor equations \eqref{eqn:ppnlintens}. Note that \(\triangle\triangle\chi\delta_{\alpha\beta}\) is a pure trace term, while \(\triangle\partial_\alpha\partial_\beta\chi\) decomposes into a pure trace and a traceless part,
\begin{equation}
\triangle\partial_\alpha\partial_\beta\chi = \triangle\triangle_{\alpha\beta}\chi + \frac{1}{3}\triangle\triangle\chi\delta_{\alpha\beta}\,,
\end{equation}
using the traceless second derivative $\triangle_{\alpha\beta}=\partial_\alpha\partial_\beta-\delta_{\alpha\beta}\triangle/3$.
In order for the tensor equations to be satisfied, both the trace and the traceless part, and thus the coefficients of both terms, must vanish,
\begin{equation}\label{eqn:36}
c^{\pm}_4 = c^{\pm}_5 = 0\,.
\end{equation}

Taking a closer look at the expansion of the coefficients displayed in appendix \ref{sec:ppnlincoeff}, one finds that the coefficients \(c^{\pm}_1, c^{\pm}_4, c^{\pm}_5\) only depend on the PPN parameters \(\alpha^{\pm}, \gamma^{\pm}, \theta^{\pm}\). We therefore have obtained the six equations \eqref{eqn:12}, \eqref{eqn:36} for six of the PPN parameters. However, due to the gauge invariance conditions \eqref{eqn:gaugeinv2}, these are linearly dependent. In order to solve the equations, one needs to (partially) fix a gauge by fixing the value of one of the parameters. The standard PPN gauge corresponds to the simple choice \(\theta^+ = 0\).

We now turn our attention to the vector equations \eqref{eqn:ppnlinvect}. The equation for \(K^{-}_{0\alpha}\) is easily solved using the fact that both the divergence-free and the total derivative part of the curvature tensor, and thus both coefficients must vanish,
\begin{equation}\label{eqn:78}
c^-_2 = c^-_3 = 0\,.
\end{equation}
Finally, we consider the equation for \(K^{+}_{0\alpha}\) which yields
\begin{equation}
c^+_2\triangle X^+_{\alpha} + c^+_3\triangle X^-_{\alpha} = -8\pi\rho v_{\alpha}
= 2\triangle V_{\alpha} = \triangle(X^+_{\alpha} + X^-_{\alpha})
\end{equation}
using the definition of $V_\alpha$ in the second equality. The equation above is now split into pure divergence and divergence-free vector terms which decouple and have to be solved independently. This results in
\begin{equation}\label{eqn:90}
c^+_2 = c^+_3 = 1\,.
\end{equation}

Another close look at the newly obtained equations for $c^\pm_3$ and the expansions given in appendix~\ref{sec:ppnlincoeff} reveals that all terms containing \(\sigma^{\pm}_-\) drop out due to the gauge invariance conditions~\eqref{eqn:gaugeinv2}. The equations for \(c^{\pm}_3\) then turn out to be linearly dependent on the equations we have obtained from the scalar and tensor components of the equations of motion, and thus they are solved identically. The remaining two equations for \(c^{\pm}_2\) can finally be used to solve for the PPN parameters~\(\sigma^{\pm}_+\).

To summarize, the linearized field equations in our multimetric PPN framework already are strong enough to determine the eight extended PPN parameters \(\alpha^{\pm}, \gamma^{\pm}, \theta^{\pm}, \sigma^{\pm}_+\). These parameters are the solutions of equations \eqref{eqn:12}, \eqref{eqn:36}, \eqref{eqn:78} and \eqref{eqn:90}. Given any particular multimetric theory consistent with our assumptions, one may use this result as a quick test of solar system consistency, simply by comparing the predicted PPN parameters with the experimentally favoured results \eqref{eqn:ppnvalues}. Before analyzing the particular theory proposed in~\cite{Hohmann:2010vt} in the following section we remark that a calculation of the remaining PPN parameters in the extended multimetric formalism requires higher order perturbation theory that also covers velocity orders $\mathcal{O}(4)$. In practice this is a very lengthy calculation that we do not wish to enter in this article, but it poses no difficulty in principle.

\section{Application to our repulsive gravity model}\label{sec:application}
We will now determine the PPN parameters \(\alpha^{\pm}, \gamma^{\pm}, \theta^{\pm}, \sigma^{\pm}_+\) of the repulsive gravity model proposed in~\cite{Hohmann:2010vt} by applying the multimetric PPN formalism developed in the previous sections. For this purpose we first derive the linearized field equations and determine the parameter matrices~\(\mat{P}, \mat{Q}, \mat{R}, \mat{M}, \mat{N}\) that appear in the linearized curvature tensor \eqref{eqn:ppngeom}. Second, we follow the steps detailed in section \ref{subsec:ppncomp} in order to compute the PPN parameters of our theory explicitly. It will turn out that these do not agree with the values obtained from experiments. But this problem can be solved, as we will finally show, by simple correction terms that improve the action of our theory. The calculated PPN parameters of the improved theory are now consistent with experiment, and the theory features the same accelerating late-time cosmology.

\subsection{Linearized field equations}\label{subsec:lineareom}
The action of our repulsive multimetric gravity model~\cite{Hohmann:2010vt} is of the form~\eqref{eqn:actionsplit}. The gravitational part of the action is explicitly given by
\begin{equation}\label{eqn:gravaction}
S_G[g^1, \ldots, g^N] = \frac{1}{16\pi}\int d^4x\sqrt{g_0}\sum_{I, J = 1}^{N}(x + y\delta^{IJ})g^{Iij}R^J_{ij}
\end{equation}
with density $g_0=\prod_{I = 1}^{N}\left(g^I\right)^{1/N}$, and the matter action is a sum of standard model actions
\begin{equation}\label{eqn:matteraction}
S_M[g^I, \Psi^I] = \int d^4x \sqrt{g^I}L_M[g^I, \Psi^I]\,.
\end{equation}
In the case \(N = 1\) when only a single metric is present, the action becomes identical to the Einstein-Hilbert action for an appropriate choice of the constant parameters \(x, y\). The field equations are obtained by variation and take the form
\begin{equation}\label{eqn:full}
K^I_{ab} = 8\pi T^I_{ab}
\end{equation}
with the geometry tensor
\begin{eqnarray}\label{eqn:fulleom}
K^I_{ab} &=& \sqrt{{g_0}/{g^I}}\Bigg[-\frac{1}{2N}g^I_{ab}\sum_{J, K = 1}^{N}(x + y\delta^{JK})g^{Jij}R^K{}_{ij} + \sum_{J = 1}^{N}(x + y\delta^{IJ})R^J{}_{ab}\nonumber\\
&&{}- \left(2\delta^d_{(a}g^I_{b)(i}\delta^c_{j)} - g^I_{ab}\delta^c_{(i}\delta^d_{j)} - g^{Icd}g^I_{i(a}g^I_{b)j}\right)\sum_{J = 1}^{N}(x + y\delta^{IJ})\bigg(2g^{Jpi}S^{IJj}{}_{p(c}\tilde{S}^I{}_{d)} \\
&&{}+\frac{1}{2}g^{Jij}\tilde{S}^I{}_c\tilde{S}^I{}_d+\frac{1}{2} g^{Jij}\nabla^I_c\tilde{S}^I{}_d+ \nabla^I_cS^{IJi}{}_{dp} g^{Jjp} + S^{IJp}{}_{cq}S^{IJi}{}_{dp}g^{Jjq} + S^{IJi}{}_{cq}S^{IJj}{}_{dp}g^{Jpq}\bigg)\Bigg],\nonumber
\end{eqnarray}
where the connection difference tensors are defined as
\begin{equation}
S^{IJi}{}_{jk} = \Gamma^{Ii}{}_{jk}-\Gamma^{Ji}{}_{jk}\,,\quad
S^{IJ}{}_j=S^{IJk}{}_{jk}\,, \quad
\tilde S^{Ji}{}_{jk}=\frac{1}{N}\sum_{I=1}^N S^{IJi}{}_{jk}\,,\quad
\tilde S^{J}{}_j=\tilde S^{Jk}{}_{jk}\,.
\end{equation}

We now derive the linearized field equations using the perturbative ansatz \(g^I_{ab} = \eta_{ab} + h^I_{ab}\). Note that the connection differences \(S^{IJi}{}_{jk}\) and the Ricci tensor \(R^I_{ij}\) are of first order in the metric perturbations \(h^I\), so any terms containing products of two connection differences drop out; covariant derivatives acting on connection differences are replaced by ordinary partial derivatives; the metric tensors \(g^I\) are replaced by the flat metric \(\eta\) whenever they appears in a product with a connection difference or Ricci tensor. These handy rules significantly simplify the computation and one finally obtains the linearized geometry tensor
\begin{equation}\label{eqn:lineom}
\begin{split}
K^I_{ab} &= \sum_{J = 1}^{N}\Bigg[\left(2x - (Nx - y)\delta^{IJ}\right)\partial^p\partial_{(a}h^J_{b)p} + \left(-x + \frac{1}{2}(Nx - y)\delta^{IJ}\right)\square h^J_{ab}\\
&\phantom{=}+ \left(\frac{Nx}{2}\delta^{IJ} - x - \frac{y}{2N}\right)\partial_a\partial_bh^J + \left(\frac{Nx}{2}\delta^{IJ} - x - \frac{y}{2N}\right)\partial_p\partial_qh^{Jpq}\eta_{ab}\\
&\phantom{=}+ \left(x + \frac{y}{N} - \frac{Nx + y}{2}\delta^{IJ}\right)\square h^J\eta_{ab}\Bigg]+\mathcal{O}(h^2)\,.
\end{split}
\end{equation}
Comparing this equation with the most general form of the linearized geometry tensor~\eqref{eqn:ppngeom} and writing the parameter matrices in the form~\eqref{eqn:parmatsym}, we read off the parameter values
\begin{subequations}\label{eqn:parmats}
\begin{align}
P^+ &= (2 - N)x + y\,, & P^- &= 2x\,,\\
Q^+ &= \frac{N - 2}{2}x - \frac{y}{2}\,, & Q^- &= -x\,,\\
R^+ &= \frac{N - 2}{2}x - \frac{y}{2N}\,, & R^- &= -x - \frac{y}{2N}\,,\\
M^+ &= \frac{N - 2}{2}x - \frac{y}{2N}\,, & M^- &= -x - \frac{y}{2N}\,,\\
N^+ &= \frac{2 - N}{2}x + \frac{2 - N}{2N}y\,, & N^- &= x + \frac{y}{N}\,.
\end{align}
\end{subequations}
These values of course satisfy the gauge-invariance conditions~\eqref{eqn:gaugeinv2}, since they result from a diffeomorphism invariant gravitational action.

Now we are in the position to follow the steps detailed in section \ref{subsec:ppncomp} to compute the PPN parameters of our theory.

\subsection{PPN parameters}
With the parameter values of the linearized theory obtained in~\eqref{eqn:parmats} we now compute the PPN parameters $\alpha^{\pm}, \gamma^{\pm}, \theta^{\pm}, \sigma^{\pm}_+$ of our repulsive gravity model defined by \eqref{eqn:gravaction} and \eqref{eqn:matteraction}. The procedure for this is based on solving the equations of motion, for which we developed the technology in section~\ref{subsec:ppncomp}. A summary of the necessary steps is given at the end of that section. Choosing the standard PPN gauge such that \(\theta^+ = 0\) yields the PPN parameters
\begin{align}\label{eqn:ppn1}
\alpha^+ &= \frac{1}{3N}\left(\frac{3}{Nx + y} - \frac{4N - 4}{Nx - y}\right)\,, & \alpha^- &= \frac{1}{3N}\frac{7Nx + y}{N^2x^2 - y^2}\,,\nonumber\\
\gamma^+ &= \frac{1}{3N}\left(\frac{3}{Nx + y} - \frac{2N - 2}{Nx - y}\right)\,, & \gamma^- &= \frac{1}{3N}\left(\frac{3}{Nx + y} - \frac{N - 2}{Nx - y} + \frac{N}{y}\right)\,,\nonumber\\
\theta^+ &= 0\,, & \theta^- &= \frac{1}{3Nx - 3y} - \frac{1}{3y}\,,\\
\sigma_+^+ &= \frac{(2N - 4)x + 2y}{N^2x^2 - y^2}\,, & \sigma_+^- &= -\frac{4x}{N^2x^2 - y^2}\,.\nonumber
\end{align}

We now focus on the Newtonian limit of our theory. Recall that we demanded in~\cite{Hohmann:2010vt} a Newtonian limit where the gravitational interaction within each sector is attractive, while it is repulsive of equal strength between matter belonging to different sectors. This limit corresponds to the PPN parameters \(\alpha^+ = 1\) and \(\alpha^- = -1\). These are achieved for parameter values
\begin{equation}
x = \frac{2N - 1}{6N(2 - N)}, \quad y = \frac{-2N + 7}{6(2 - N)}\,.
\end{equation}
Note that this recovers the same values that can also be obtained from a purely Newtonian calculation~\cite{Hohmann:2010vt}. Substituting these values into~\eqref{eqn:ppn1} simplifies the results for the PPN parameters to
\begin{align}
\alpha^+ &= 1\,, & \alpha^- &= -1\,,\nonumber\\
\gamma^+ &= \frac{1}{N}\,, & \gamma^- &= \frac{3}{2N - 7} + \frac{1}{N} + \frac{1}{2}\,,\nonumber\\
\theta^+ &= 0\,, & \theta^- &= \frac{1}{7 - 2N} - \frac{3}{2}\,,\\
\sigma_+^+ &= -1 - \frac{1}{N}\,, & \sigma_+^- &= 2 - \frac{1}{N}\,.\nonumber
\end{align}
Comparison with the observed values \(\gamma^+ = 1\) and \(\sigma^+_+ = -2\) displayed in~\eqref{eqn:ppnvalues} immediately shows that these are satisfied only in the case \(N = 1\), i.e., when there is only one metric and a corresponding copy of the standard model, in which case our theory reduces to Einstein gravity. This is a dissatisfactory result since our aim was the construction of experimentally feasible gravity theories for \(N > 1\).

This result shows that our model requires modification in order to match experimental bounds from solar system experiments. In the following we will make such improvements that adapt the theory to the observed values of the PPN parameters.

\subsection{Improved PPN consistent model}\label{subsec:extension}
Since the PPN parameters calculated for our theory do not match the observed values, it is natural to ask whether the theory can be modified so to reproduce the correct values. We will now show that this is indeed possible.

We will modify the action and then proceed in complete analogy to the previous sections. First, we will compute the field equations from a variation of the modified action. Second, we will expand the metric around a flat solution and derive the linearized field equations. Third, we will read off the values of the parameter matrices and employ the linearized multimetric PPN formalism constructed in section~\ref{subsec:ppncomp}. We will only give a brief sketch of this calculation here.

We start from the gravitational action \eqref{eqn:gravaction} and add the following term which is consistent with our assumptions {\it (i)}--{\it (iv)} of section \ref{sec:multippn} that restrict the multimetric theories we consider in this article:
\begin{equation}\label{eqn:acmod}
\tilde{S}_G = \frac{1}{16\pi}\sum_{I = 1}^N\int d^4x\sqrt{g_0}\,g^{I\,ij}\left(z\tilde{S}^I{}_k\tilde{S}^{I\,k}{}_{ij} + u\tilde{S}^I{}_i\tilde{S}^I{}_j\right).
\end{equation}
This term contains two new constant parameters \(z, u\) that will be determined by PPN consistency below. The above modification is not the only possibility to achieve experimental consistency of our model, as we will discuss in the conclusion.

Here, we continue by computing the equations of motion by variation. These take the form displayed in equations~\eqref{eqn:full}, but the curvature tensor \(K^I_{ab}\) in~\eqref{eqn:fulleom} attains a correction term $\tilde{K}^I_{ab}$ that is rather involved. We then compute the linearized curvature tensor. In addition to the result obtained in~\eqref{eqn:lineom} this gives
\begin{equation}\label{eqn:lincorr}
\begin{split}
\tilde{K}^I_{ab} &= \frac{z}{2}\eta_{ab}\eta^{ij}\partial_k\tilde{S}^{I\,k}{}_{ij} + \frac{2u - z}{2}\eta_{ab}\eta^{ij}\partial_i\tilde{S}^I{}_j + z\partial_{(a}\tilde{S}^I{}_{b)}+\mathcal{O}(h^2)\\
&= \sum_J\left(-\frac{1}{N} + \delta^{IJ}\right)\!\left(\frac{z}{2}\partial_a\partial_bh^J + \frac{z}{2} \partial_p\partial_qh^{J\,pq}\eta_{ab} + \frac{z - u}{2}\square h^J\eta_{ab}\right)+\mathcal{O}(h^2)\,.
\end{split}
\end{equation}
Next, we read off the modified parameter matrices \(\mat{P}, \mat{Q}, \mat{R}, \mat{M}, \mat{N}\). Using the notation introduced in section~\ref{subsec:geometry}, we obtain the following modified results as compared to~\eqref{eqn:parmats}:
\begin{subequations}
\begin{align}
P^+ &= (2 - N)x + y\,, & P^- &= 2x\,,\\
Q^+ &= \frac{N - 2}{2}x - \frac{y}{2}\,, & Q^- &= -x\,,\\
R^+ &= \frac{N - 2}{2}x - \frac{y}{2N} + \frac{N - 1}{2N}z\,, & R^- &= -x - \frac{y}{2N} - \frac{z}{2N}\,,\\
M^+ &= \frac{N - 2}{2}x - \frac{y}{2N} + \frac{N - 1}{2N}z\,, & M^- &= -x - \frac{y}{2N} - \frac{z}{2N}\,,\\
N^+ &= \frac{2 - N}{2}x + \frac{2 - N}{2N}y + \frac{N - 1}{2N}(z - u)\,, & N^- &= x + \frac{y}{N} - \frac{z - u}{2N}\,.
\end{align}
\end{subequations}

Finally, we follow the steps detailed in section~\ref{subsec:ppncomp} to compute the PPN parameters \(\alpha^{\pm}, \gamma^{\pm}, \theta^{\pm}, \sigma^{\pm}_+\). In comparison to~\eqref{eqn:ppn1} these take the modified values
\begin{gather}
\alpha^+ = \frac{1}{3N}\left(\frac{3}{Nx + y} - \frac{4N - 4}{Nx - y}-\frac{2(N-1)(u-3z)}{\Xi}\right),\nonumber\\
\alpha^- = \frac{1}{3N}\left(\frac{7Nx + y}{N^2x^2 - y^2}+\frac{2(u-3z)}{\Xi}\right),\nonumber\\
\gamma^+ = \frac{1}{3N}\left(\frac{3}{Nx + y} - \frac{2N - 2}{Nx - y}+\frac{2(N-1)(u-3z)}{\Xi}\right),\nonumber\\
\gamma^- = \frac{1}{3N}\left(\frac{3}{Nx + y} - \frac{N - 2}{Nx - y} + \frac{N}{y}+\frac{(4N-2)u+(6-9N)z+3Ny}{\Xi}\right),\\
\theta^+ = 0\,,\qquad \theta^- = \frac{1}{3Nx - 3y} - \frac{1}{3y}-\frac{4u+3(y-3z)}{3\Xi}\,,\nonumber\\
\sigma_+^+ = \frac{(2N - 4)x + 2y}{N^2x^2 - y^2}\,,\qquad \sigma_+^- = -\frac{4x}{N^2x^2 - y^2}\,.\nonumber
\end{gather}
for $\Xi=3(y^2+z^2)-2Nx(u-3z)+2uy$. One can now choose the parameters \(x, y, z, u\) so that not only the Newtonian limit agrees with our repulsive gravity requirement \(\alpha^+ = -\alpha^- = 1\), but also the observed PPN parameter values \(\gamma^+ = 1\) and \(\sigma^+_+ = -2\) are obtained. These results are achieved for parameter values
\begin{equation}
x = \frac{1}{8 - 4N}\,, \quad y = \frac{4 - N}{8 - 4N}\,, \quad z = -\frac{4 - N}{8 - 4N}\,, \quad u = -\frac{12 - 3N}{8 - 4N}\,.
\end{equation}
The remaining PPN parameters then are determined to be \(\gamma^- = -1\), \(\theta^- = 0\) and \(\sigma^-_+ = 2\).

Now that the PPN parameters of the improved model are consistent as far as we can determine from the linearized multimetric PPN formalism, one may ask whether the remarkable cosmological features of the original theory presented in~\cite{Hohmann:2010vt}, such as the accelerating late time expansion and the big bounce, are still present in the improved version.

It can easily be seen that this is the case by noting that the modification~\eqref{eqn:acmod} of the action is quadratic in the connection difference tensors \(S^{IJ}\). Consequently the additional terms in the equations of motion caused by this modification contain at least one connection difference tensor. The remains of these are also seen in the linearized term~\eqref{eqn:lincorr}. For our simple cosmological model we assumed that a cosmological version of the Copernican principle holds in the sense that equal amounts of each type of matter are distributed homogeneously in our universe, and thus we could argue that all metric tensors of the cosmological solution should be equal at very early and very late times. It then follows that the connection differences vanish and the earlier obtained cosmological equations of motion are unchanged under the modifications we presented in this section.

\section{Conclusion}\label{sec:conclusion}
This article continued our discussion of multimetric gravity theories in which the gravitational field is described by $N$ metrics, and which contain a corresponding number of standard model copies~\cite{Hohmann:2009bi,Hohmann:2010vt}. The motivation to study these theories comes from the fact that they can be constructed so that the astronomy of $N-1$ matter sectors appears dark for any given observer, and (for $N>2$) so that different types of matter feature a mutual gravitational repulsion in the Newtonian limit. In this way they may naturally model cosmological effects such as late-time acceleration.

From the particle theorist's point of view it can be regarded as a strength of our models that they do not introduce matter fields of unknown masses, charges or couplings. The non-gravitational particle content of the well-understood standard model is simply copied into the different sectors. Interactions between the sectors are mediated only through the coupling of the different metrics, as becomes clear from the action structure (\ref{eqn:actionsplit}); the relevant cross sections will involve the Newton's constant squared. Hence, direct experimental observation of the other matter types will be extremely difficult.

To discuss the gravitational field content of our models, we repeat an observation from~\cite{Hohmann:2009bi}: while $N$ symmetric two-tensors $h^I_{ab}$ appear on the linearized level, diffeomorphism-invariance merely implies a single gauge symmetry under $\delta_\xi h^I_{ab} = 2 \partial_{(a}\xi_{b)}$ for common gauge parameters $\xi_b$. This type of gauge symmetry is required for the definition of a massless particle of spin two~\cite{vanderBij:1981ym}. Since every observer in our theories can choose to relate the gauge symmetry to his own metric field, we may interpret our theories as containing one graviton and further $N-1$ symmetric two tensor fields that cannot be interpreted as spin two particles~\cite{vanderBij:1981ym}.

In this article we asked whether theories of this type are consistent with data available from high precision experiments at the solar system level. A suitable framework to test gravity theories based on a single metric exists, and is known as the parametrized post-Newtonian formalism. In this article we have constructed a simple extension of this PPN formalism to multimetric gravity theories. We worked on the assumption that regions exist which are dominated by matter belonging to a single standard model copy, and where one may neglect the influence of matter from different sectors on the gravitational field. This is a reasonable assumption for repulsive gravity theories of our type because different types of matter should separate as the universe evolves.

Our multimetric extension of the PPN formalism features an additional set of PPN parameters describing the mutual gravitational interaction of matter belonging to different sectors. We have shown that a subset of these parameters can be obtained already from the linearized equations of motion. This yields a quick method of selecting consistent theories or discarding unphysical ones. We have applied the linearized multimetric PPN formalism to the particular gravity theory presented in~\cite{Hohmann:2010vt}. We computed the PPN parameters and found that these do not agree with the observed values. We finally could present an improved theory which now does agree with observations while still featuring the same promising cosmological effects as the original theory, most notably the accelerating late-time expansion of the universe.

It is worth pointing out that the particular improvement~\eqref{eqn:acmod} of our multimetric repulsive gravity model is not unique. Two new parameters control the modification terms and need to be determined by experimental consistency requirements. Even within the class of quadratic connection difference terms, one could discuss other examples of modification, with even more additional parameters, that also yield correct PPN values without changing the cosmological properties. The linearized multimetric PPN formalism developed in this article does not well distinguish theories of this type.

Hence it would be desirable to find and establish further physical and mathematical principles in order to restrict the possible terms in multimetric gravity actions. The most obvious restriction one should pose is that the remaining PPN parameters that cannot be obtained from the linearized theory should also agree with the observed values. The determination of these parameters would require a perturbative expansion of the equations of motion up to quadratic order in the metric perturbations which is very involved for multimetric gravity theories, but should be carried out in future work. One mathematical idea to restrict possible gravitational actions could be to enlarge the symmetry group. While we restricted to a discrete exchange symmetry with respect to arbitrary permutations of the sectors, one could think of establishing a continuous symmetry group that mixes the sectors. In any case, the search for further principles behind multimetric gravity actions is still an open question.

\appendix
\section{Coefficients of the linearized PPN ansatz}\label{sec:ppnlincoeff}
In this appendix we display the detailed expression for the coefficients \(c^{\pm}_1, \ldots, c^{\pm}_5\) used in the expansion of the geometry tensor given in equation \eqref{eqn:ppnlincoeff}. These can be computed using the expression for the linearized geometry tensor \eqref{eqn:ppngeom} and the linearized PPN metric ansatz \eqref{eqn:ppnlinmet}:
%\begin{subequations}
\begin{equation}
\begin{split}
c^+_1 &= -(N^+ + Q^+)\alpha^+ - (N - 1)(N^- + Q^-)\alpha^-\\
&\phantom{=}+ (M^+ + 3N^+)\gamma^+ + (N - 1)(M^- + 3N^-)\gamma^-\\
&\phantom{=}- (M^+ - N^+)\theta^+ - (N - 1)(M^- - N^-)\theta^-\,,
\end{split}
\end{equation}
\begin{equation}
\begin{split}
c^-_1 &= -(N^- + Q^-)\alpha^+ - (N^+ + Q^+ + (N - 2)(N^- + Q^-))\alpha^-\\
&\phantom{=}+ (M^- + 3N^-)\gamma^+ + (M^+ + 3N^+ + (N - 2)(M^- + 3N^-))\gamma^-\\
&\phantom{=}- (M^- - N^-)\theta^+ - (M^+ - N^+ + (N - 2)(M^- - N^-))\theta^-\,,
\end{split}
\end{equation}
\begin{equation}
c^+_2 = Q^+\sigma^+_+ + (N - 1)Q^-\sigma^-_+\,,
\end{equation}
\begin{equation}
c^-_2 = Q^-\sigma^+_+ + (Q^+ + (N - 2)Q^-)\sigma^-_+\,,
\end{equation}
\begin{equation}
\begin{split}
2c^+_3 &= (P^+ + 2R^+)\alpha^+ + (N - 1)(P^- + 2R^-)\alpha^-\\
&\phantom{=}- (P^+ + 6R^+)\gamma^+ - (N - 1)(P^- + 6R^-)\gamma^-\\
&\phantom{=}+ (P^+ - 2R^+)\theta^+ + (N - 1)(P^- - 2R^-)\theta^-\\
&\phantom{=}+ (P^+ + 2Q^+)\sigma^+_- + (N - 1)(P^- + 2Q^-)\sigma^-_-\,,
\end{split}
\end{equation}
\begin{equation}
\begin{split}
2c^-_3 &= (P^- + 2R^-)\alpha^+ + (P^+ + 2R^+ + (N - 2)(P^- + 2R^-))\alpha^-\\
&\phantom{=}- (P^- + 6R^-)\gamma^+ - (P^+ + 6R^+ + (N - 2)(P^- + 6R^-))\gamma^-\\
&\phantom{=}+ (P^- - 2R^-)\theta^+ + (P^+ - 2R^+ + (N - 2)(P^- - 2R^-))\theta^-\\
&\phantom{=}+ (P^- + 2Q^-)\sigma^+_- + (P^+ + 2Q^+ + (N - 2)(P^- + 2Q^-))\sigma^-_-\,,
\end{split}
\end{equation}
\begin{equation}
\begin{split}
c^+_4 &= N^+\alpha^+ + (N - 1)N^-\alpha^-\\
&\phantom{=}- (M^+ + 3N^+ + Q^+)\gamma^+ - (N - 1)(M^- + 3N^- + Q^-)\gamma^-\\
&\phantom{=}+ (M^+ - N^+ - Q^+)\theta^+ + (N - 1)(M^- - N^- - Q^-)\theta^-\,,
\end{split}
\end{equation}
\begin{equation}
\begin{split}
c^-_4 &= N^-\alpha^+ + (N^+ + (N - 2)N^-)\alpha^-\\
&\phantom{=}- (M^- + 3N^- + Q^-)\gamma^+ - (M^+ + 3N^+ + Q^+ + (N - 2)(M^- + 3N^- + Q^-))\gamma^-\\
&\phantom{=}+ (M^- - N^- - Q^-)\theta^+ + (M^+ - N^+ - Q^+ + (N - 2)(M^- - N^- - Q^-))\theta^-\,,
\end{split}
\end{equation}
\begin{equation}
\begin{split}
c^+_5 &= R^+\alpha^+ + (N - 1)R^-\alpha^-\\
&\phantom{=}- (P^+ + 3R^+)\gamma^+ - (N - 1)(P^- + 3R^-)\gamma^-\\
&\phantom{=}+ (P^+ + 2Q^+ - R^+)\theta^+ + (N - 1)(P^- + 2Q^- - R^-)\theta^-\,,
\end{split}
\end{equation}
\begin{equation}
\begin{split}
c^-_5 &= R^-\alpha^+ + (R^+ + (N - 2)R^-)\alpha^-\\
&\phantom{=}- (P^- + 3R^-)\gamma^+ - (P^+ + 3R^+ + (N - 2)(P^- + 3R^-))\gamma^-\\
&\phantom{=}+ (P^- + 2Q^- - R^-)\theta^+ + (P^+ + 2Q^+ - R^+ + (N - 2)(P^- + 2Q^- - R^-))\theta^-\,.
\end{split}
\end{equation}
%\end{subequations}

\acknowledgments
MH and MNRW gratefully acknowledge full financial support from the German Research Foundation DFG through the Emmy Noether excellence grant WO 1447/1-1.

%%%%%%%%%%%%%%%%%%%%%%%%%%%%%%%%%%%%%%%%%%%

\end{document}